\documentclass[prb, showpacs, preprintnumbers, amsmath, amssymb, superscriptaddress, twocolumn]{revtex4}
\usepackage{graphicx, color, dcolumn, bm, natbib, amssymb, amsmath,color}

\newcommand{\m}{\boldsymbol}
\usepackage[final=true]{hyperref}
\usepackage[normalem]{ulem}
\begin{document}
\title{Spin-wave dynamics in FeGe helimagnet: \\ studied by small-angle neutron scattering}
\author{S.-A. Siegfried}
\affiliation{German Engineering Materials Science Centre (GEMS) at Heinz Maier-Leibnitz Zentrum (MLZ), Helmholtz-Zentrum Geesthacht GmbH, Lichtenbergstr. 1, 85747 Garching bei M\"unchen, Germany}
\author{A. S. Sukhanov}
\affiliation{Petersburg Nuclear Physics Institute, Gatchina, St-Petersburg, 188300, Russia}
\affiliation{Saint-Petersburg State University, Ulyanovskaya 1, Saint-Petersburg, 198504, Russia}
\author{E. V. Altynbaev}
\affiliation{Petersburg Nuclear Physics Institute, Gatchina, St-Petersburg, 188300, Russia}
\affiliation{Saint-Petersburg State University, Ulyanovskaya 1, Saint-Petersburg, 198504, Russia}
\author{D. Honnecker}
\affiliation{Institute Laue Langevin, Grenoble, 38042 Grenoble, Cedex 9, France}
\author{A. Heinemann}
\affiliation{German Engineering Materials Science Centre (GEMS) at Heinz Maier-Leibnitz Zentrum (MLZ), Helmholtz-Zentrum Geesthacht GmbH, Lichtenbergstr. 1, 85747 Garching bei M\"unchen, Germany}
\author{A. V. Tsvyashchenko}
\affiliation{Institute for High Pressure Physics, Russian Academy of Sciences, 142190 Troitsk, Moscow, Russia}
\author{S. V. Grigoriev}
\affiliation{Petersburg Nuclear Physics Institute, Gatchina, St-Petersburg, 188300, Russia}
\affiliation{Saint-Petersburg State University, Ulyanovskaya 1, Saint-Petersburg, 198504, Russia}
\date{\today}
\begin{abstract}
We have studied the spin-wave stiffness of the Dzyaloshinskii-Moriya helimagnet FeGe in a temperature range from 225~K up to $T_C \approx$~278.7~K by small-angle neutron scattering. The method we have used is based on  [S. V. Grigoriev et al. Phys. Rev. B \textbf{92} 220415(R) (2015)] and was extended here for the application in polycrystalline samples. We confirm the validity of the anisotropic spin-wave dispersion for FeGe caused by the Dzyaloshinskii-Moriya interaction. We have shown that the spin-wave stiffness $A$ for FeGe helimagnet decreases with a temperature as $A(T) = 194(1-0.7(T/T_C)^{4.2})$ meV\AA$^2$. The finite value of the spin-wave stiffness $A = 58$ meV\AA$^2$  at $T_C$ classifies the order-disorder phase transition in FeGe as being the first order one.

\end{abstract}
\maketitle
\section{Introduction}

The cubic B20 compounds have a noncentrosymmetric crystal structure described by the P2$_1$3 space group. The lack of a symmetry center of the crystal structure produces the chiral spin-spin Dzyaloshinskii-Moriya (DM) interaction \cite{Dzyaloshinskii64ZETF,Moriya1960}. 
According to the model by Bak and Jensen \cite{BakJPCM80_v13} and Kataoka \cite{Kataoka1980}, the major ferromagnetic exchange interaction $J$, together with the DM interaction $D$ produces a (homochiral) structure in these systems below $T_c$. The energy landscape in these systems is given by $J$ and $D$, which are balanced via the helix wave vector $k_s = D/J$. The anisotropic exchange interaction has been added to the model, changing slightly the value and fixing the direction of the wave vector $\boldsymbol{k}_s$ along the principle cubic axis. As noticed by Kataoka and co-workers \cite{Kataoka1980} and Maleyev and co-workers \cite{GrigorievPRB2015}, the cubic anisotropy can play an important role in the case of relative small values of the helix wave vector $k_s$. If the anisotropic energy getting comparable to the DM interaction, it can destabilize the entire helix structure and stabilizes instead the ferromagnetic state. 

The external magnetic field $H_{c1}$ is needed to rotate the helix wave vector $\boldsymbol{k}_s$ towards the field direction and, therefore, it is a measure for the anisotropy of the system. According to \cite{Maleyev_PRB_2006}, the energy difference between the conical and the collinear full-polarized (FP) state can be directly measured by the second critical field $H_{c2}$. This energy is equal to $g \mu_B H_{c2} \approx A k_{s}^{2}$, where $A = J \cdot S \cdot a^{2}$ is the spin-wave stiffness, $S$ is the ordered spin, and $a$ is the lattice constant. The experimental parameters $k_s$, $H_{c1}$, $H_{c2}$ and $S$ describe completely the magnetic system of such compounds. 

The compound FeGe shows a significant difference in the parameters of the magnetic structure compared to the other B20 helimagnet MnSi\cite{Lebech_PCM_1989,Ishikawa_SSC_1976}. The helix length is nearly four times higher and equal to $\lambda_h \approx 700$~\AA , and the ordering temperature $T_C \approx$~278.7~K is nearly ten times higher than the $T_C$ for MnSi. As to the anisotropic interactions the helix wave vector is pinned along the $\left[111\right]$ direction in MnSi at all temperatures below $T_C \approx$~29.5~K, whereas the helix wave vector in FeGe is pinned along the $\left[100\right]$ direction in the high-temperature range between $T_C$~=~278.7~K and $T_{2\downarrow}$=211~K/$T_{2\uparrow}$=245~K and rotates towards the $\left[111\right]$ direction in the low-temperature part of the ordered phase for $T < T_{2\downarrow,\uparrow}$. According to the Bak and Jensen model a temperature driven rotation of the spiral from the $\left[100\right]$ to the $\left[111\right]$ goes along with a change of the sign of the second order gradient in the free energy expansion. Nevertheless, magnetization measurements indicated that the rotation of the helix axis at $T_{2\downarrow,\uparrow}$ can be explained by an interplay of constants of 4th and 6th terms of the cubic anisotropy \cite{Lundgren_1970}.

While the spin-wave dynamic of MnSi has been intensively studied in the past \cite{Ishikawa_PRB_1977,Tarvin_PRB_1978,Semandi_PhysicaB_1999,Janoschek_PRB_2010,Kugler_PRL_2015,Schwarze_Nature_2015,GrigorievPRB2015(R)}, it was never the case for FeGe, due to the lack of a sufficient large amount of single crystalline samples, which is necessary for the triple axis spectroscopy (TAS). The recently proposed method \cite{GrigorievPRB2015(R)} to determine the spin-wave stiffness in the helical magnets based on DM interaction in the FP state, using polarized small-angle neutron scattering (SANS), can be extended to be used for the polycrystalline FeGe compounds. This method was originally developed and applied to measure the spin-wave stiffness in ferromagnetic compounds \cite{Okorov_JETP_1986,Deriglazov_PhysicaB_1992,Toperverg_PhysicaB_1993,Grigoriev_AppPhysA_2002,Grigoriev_JSI_2014}. 
The presence of the DM chiral interaction leads to the chirality of the spin-waves in FP state of helimagnets. This fact is related to the completely anisotropic dispersion relation of magnons, which reads 
\begin{equation} \label{eq:0}
	\epsilon_{\m {q}} = A\left({\m q}-{\m k}_s\right)^2 + H - H_{c2}
\end{equation}
 for the magnetic field above the critical value $H_{c2}$ \cite{Kataoka_SW}. It can be analytically shown that the inelastic scattering of the neutrons in this case is concentrated mostly around the momentum transfers corresponding to $\pm {\m k}_s$ within two narrow cones limited by the cut-off angle $\theta_C$ for the energy gain/energy loss, respectively \cite{GrigorievPRB2015(R)}. The cut-off angle is connected to the spin-wave stiffness via the dimensionless parameter $\theta_0 = \left(2Am_n\right)^{-1}$:
\begin{align} \label{eq:1}
\theta_{C}^{2}\left(H\right) = \theta_{0}^{2}- \frac{\theta_{0}}{E_i}H + \theta_{B}^{2},
\end{align}
where $m_n$ is the neutron mass, $\theta_B$ is the Bragg angle of the scattering on the spin spiral with the length $2\pi/k_s$, and $E_i$ denotes the energy of the incident neutrons.

Here we assume conditions $\omega \ll E_i$ and $\omega \ll T$ to be fulfilled in the limit of small-angle neutron scattering. The first condition allows one to split momentum transfer into elastic component perpendicular to ${\m k}_i$ and inelastic one parallel to ${\m k}_i$. Bose factor $\left[1-\exp(-\omega/T)\right]^{-1}$ can be replaced by $T/\omega$ accounting for the second condition. Despite of being $\omega$-odd, the cross section of the inelastic scattering being integrated over the energy transfer contains the polarization-dependent part due to the peculiarity of the aforementioned asymmetric dispersion relation. As was demonstrated in \cite{GrigorievPRB2015(R)}, one can distinguish the scattering from the helimagnons in a homochiral crystalline sample using polarized neutrons in SANS experiment.

\begin{figure}[h]
	{\centering
		\begin{minipage}{0.99\linewidth}
				\center{\includegraphics[width=0.85\linewidth]{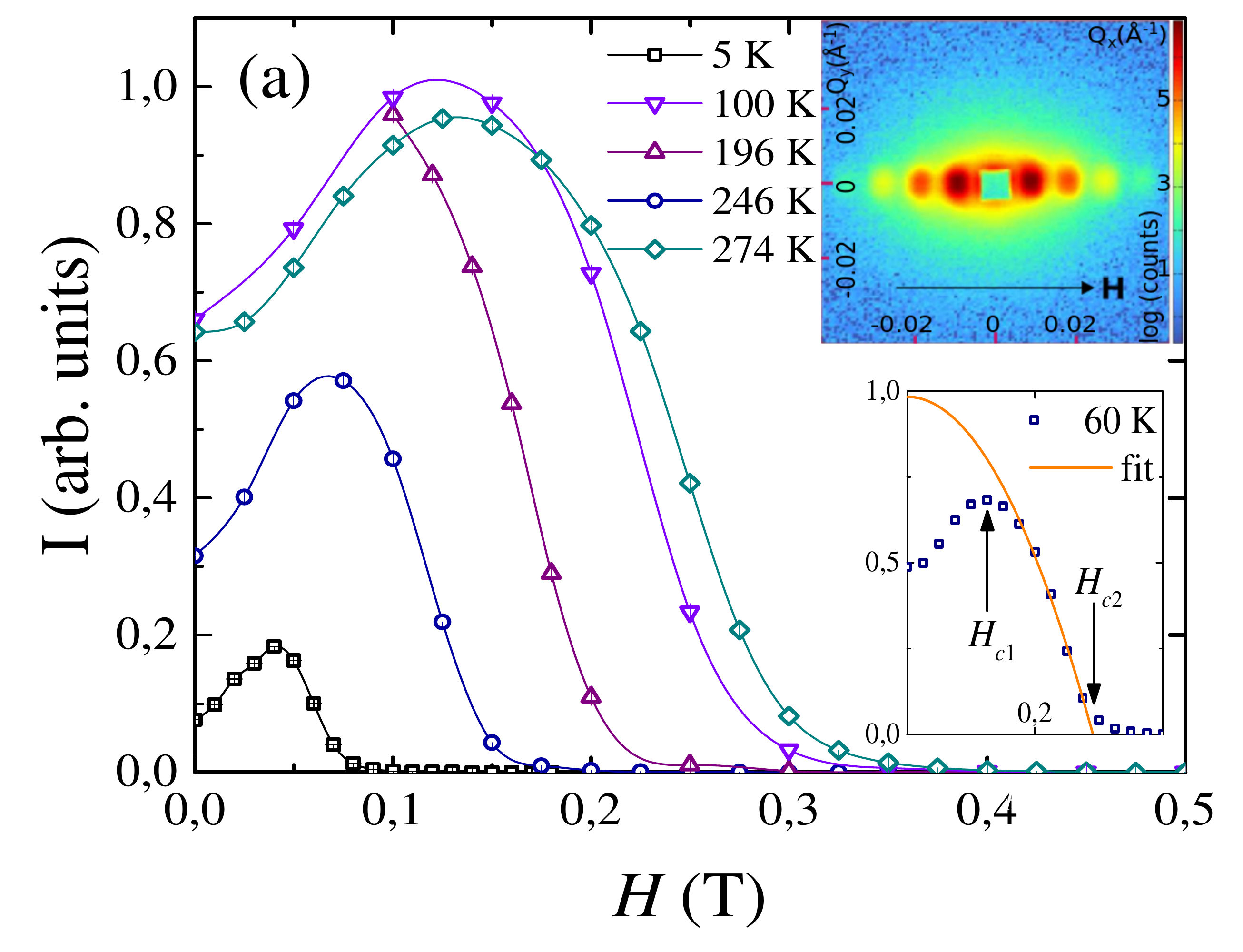}
				}
			\center{\includegraphics[width=0.99\linewidth]{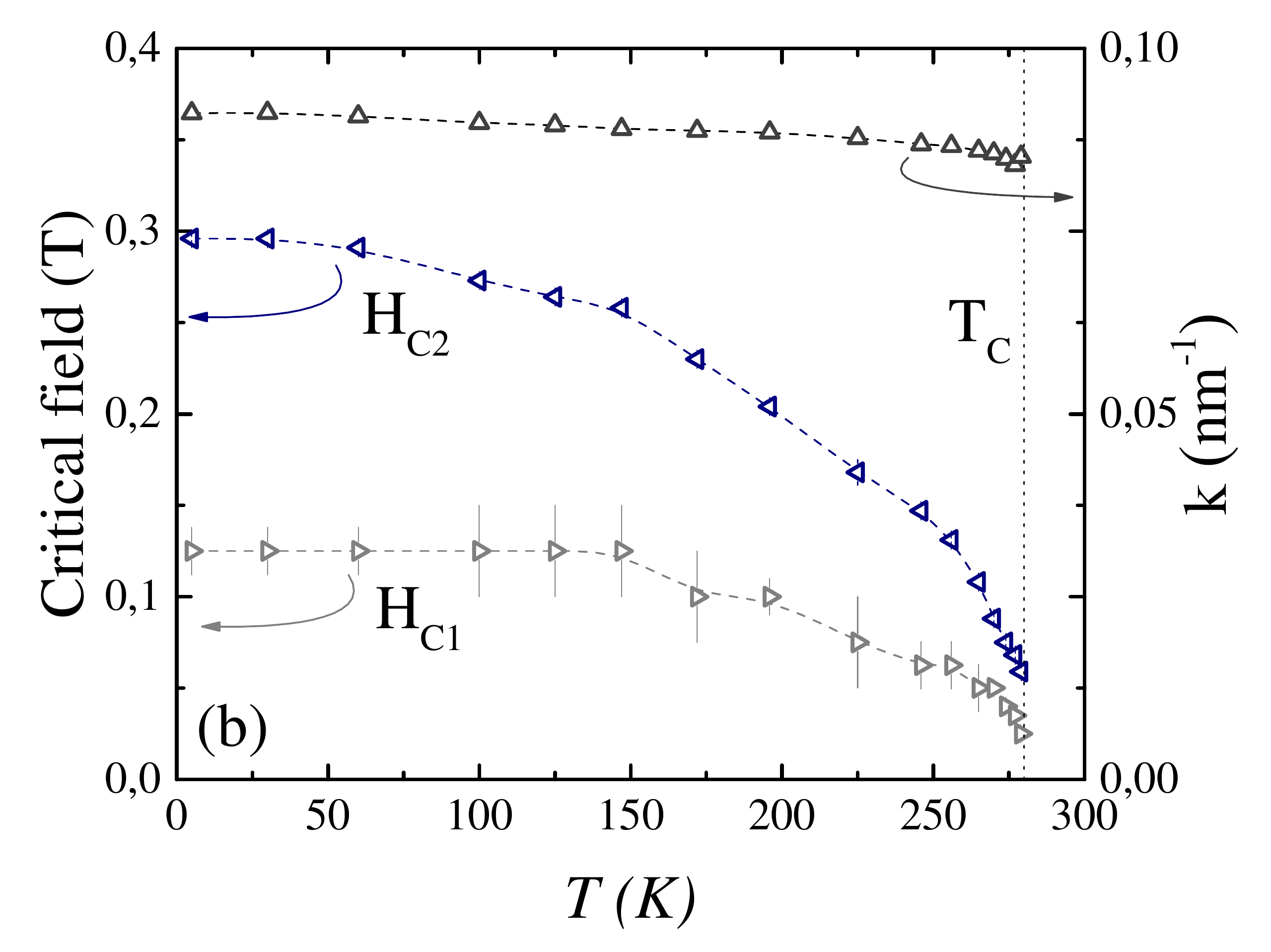}
			}
		\end{minipage}
		\caption{(color online). (a) Typical SANS scattering pattern from a monodomain helix structure below $H_{c2}$, $T =$~250~K and a magnetic field $H =$~0.075~T.  Integral intensity of the Bragg peak against the magnetic field at different temperatures between 5 and 274~K. Solid lines are guide for eyes. The upper insert shows a typical SANS scattering pattern from a monodomain helix structure below $H_{c2}$, $T =$~250~K and a magnetic field $H =$~0.075~T. The bottom insert shows an illustration of the determination of the $H_{c2}$ at 60 K. The BJ model is used to provide the fitting function (see the text for more details). (b) The temperature dependence of the value of the helix wavevector $k_s$ and both critical field $H_{c1,2}$. Dashed lines are guide for eyes.}
		\label{ris:fig1}}
\end{figure}

In this paper we show the possibility to measure the SW stiffness in the polycrystalline samples using non-polarized neutrons.  As the sample contains both left and right crystallites, it is, therefore, not possible to demonstrate the chiral nature of the spin-wave scattering. Nevertheless, it is possible to detect two circles on a scattering map with radius $\theta_C$ and centered at the Bragg peak $\theta_B$. The spin-wave stiffness $A$ is measured in the temperature range below $T_C$ by finding the cut-off angle in accordance to Eq.\ref{eq:1}. 

This paper is organized in the following way: Sections \ref{sec:II} and \ref{sec:III} give the results of the small-angle neutron scattering measurements of the FeGe compound. Section \ref{sec:IV} presents the conclusions.

\section{Elastic small-angle neutron measurements}\label{sec:II}
The SANS experiment was performed at the D11 instrument at the ILL (Grenoble, France). An unpolarized beam with a mean wavelength of $\lambda = 0.6$~nm was used. A magnetic field (0.0 - 0.5~T) was applied perpendicular to the incident beam. The FeGe sample with the mass of 0.1 g was the same as used in our previous work \cite{Grigoriev_PRB_2014}. It was synthesized using the high pressure method (see \cite{Tsvyashchenko_1984} for details).

Typical SANS map for $T =$~250~K and $H \gtrsim H_{c1} = $~0.075~T is shown in Fig.\ref{ris:fig1} (a). Several Bragg peaks appear on the left and right sides of the scattering map, where the center of the map corresponds to $Q_x = 0, Q_y = 0$. The two peaks closest to the center are from the helical structure at $Q = \pm 0.09$ nm$^{-1}$. The others are clearly the higher orders of multiple scattering. The spiral wave vector is aligned to the direction of the magnetic field. 

Fig.\ref{ris:fig1} (b) represents the integrated intensities of the Bragg peak $I$ at $Q = \pm 0.09$ nm$^{-1}$ as function of the field for different temperatures between 5 K and $T_C \approx$~278.7~K. The sample has been cooled down from $T_C$. The cooling has been stopped at 16 different temperatures in-between and a field scan with increasing magnetic field was performed at each  temperature. As it is shown in the insert, the first critical field $H_{c1}$ is defined as a point of the maximal intensity, this is the point where all spirals are aligned along the field. The difference to the previously used method \cite{Grigoriev_PRB_2014} for Fe$_{1-x}$Co$_x$Ge is worth mentioning, where we determined $H_{c1}$ as the starting point for the increasing intensity, corresponding to the spirals starts aligning along the field direction. The second critical field $H_{c2}$ is determined from the zero point of the fitting function according to the Bak-Jensen model. The model predicts that the cone angle $\alpha$, counted from the helix plane, increases with the field as \cite{Maleyev_PRB_2006}:
\begin{equation}
\sin\alpha = \frac{H}{H_{c2}}.
\end{equation}
Thus, the intensity of elastic scattering subsides $I \sim \cos^2\alpha \sim 1 - \left(H/H_{c2}\right)^2$. However, a tail of the intensity can be still observed above $H_{c2}$ (Fig. \ref{ris:fig1} (a)). This tail is better pronounced at the low temperatures, while it becomes invisible in the high-temperature region. The phenomenon may be caused by influence of the cubic anisotropy that makes the critical field $H_{c2}$ depending on orientation of the applied field with respect to the principal crystallographic axes \cite{GrigorievPRB2015}. The magnetic structure in the randomly oriented crystallites undergo the phase transition to the field polarized state at different strength of the field from the minimal $H_{c2} = Ak_s^2 - 8G/(3S)$ to maximal value $H_{c2} = Ak_s^2 + 4G/S$, where $G$ is the constant of the cubic anisotropy.

The helix wave vector $| \boldsymbol{k}_s|$ has been determined as a center of the Gaussian function fitting the helical Bragg peak in zero magnetic field for each temperature.
The temperature dependence of the helix wave vector is shown in Fig.\ref{ris:fig1} (b), it is nearly constant and equal to 0.09~nm$^{-1}$ in the whole temperature range. Furthermore, the $H-T$ phase diagram for FeGe between 5~K and $T_C$ is shown in Fig.\ref{ris:fig1} (b). The second critical field $H_{c2}$ decreases slowly from a maximum value of 0.3~T at low temperature and tends to zero at $T_C$. The first critical field $H_{c1}$ stays roughly constant at 0.1~T between 5~K and 180~K and decreases with further temperature increase towards zero at $T_C$. As it was mentioned, the $H_{c2}$ is the measure of the difference in energies of the spiral state and the FP state and is equal to $Ak_s^2$.  As far as $k_s$ shows no change, temperature decrease of $H_{c2}$ is expected to be driven by decrease of $A$ and is related to the softening of the magnetic structure with the temperature. Meanwhile the mechanism standing behind the temperature changing of the $H_{c1}$ remains less clear. 
\begin{figure}[h]
	\begin{minipage}{0.99\linewidth}
		\center{\includegraphics[width=1\linewidth]{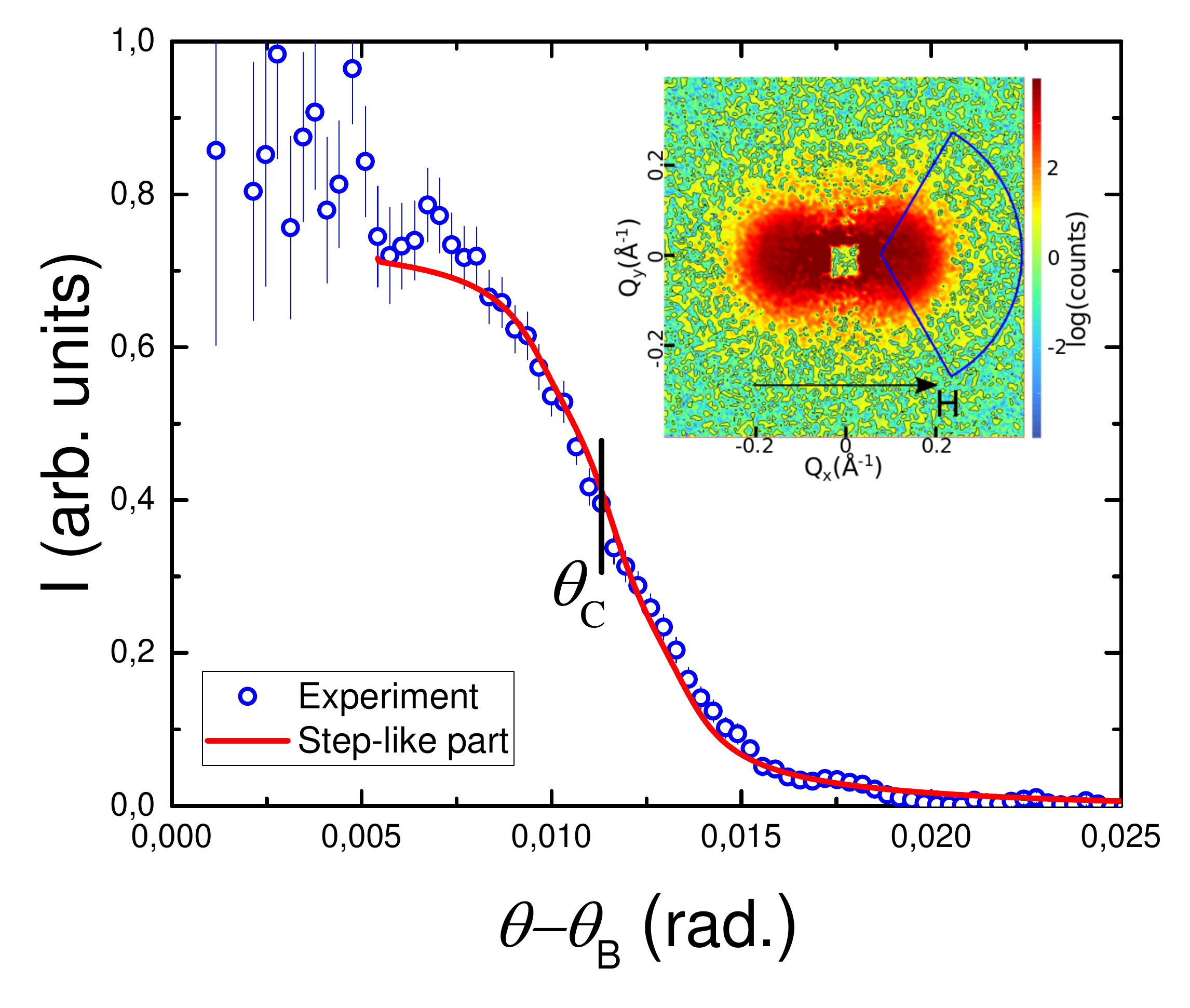}}
	\end{minipage}
	\caption{(color online).The averaged intensity profile at $T = 250$ K and $H = 0.3$ T. Sigmoid function fits the step-like part of the scattering. The cut-off angle is shown. Inset: SANS map above the critical field, $T = 250$ K and $H = 0.3$ T.}
	\label{ris:fig2}
\end{figure}

\section{Small-angle neutron measurements  of the spin-wave stiffness}\label{sec:III}

The insert of Figure \ref{ris:fig2} shows a typical SANS map for FeGe, which is taken above $H_{c2}$. As the field reaches $H_{c2}$ the elastic scattering disappears and only the inelastic scattering centered at $\boldsymbol{Q} = \pm \boldsymbol{k}_s$ remains. This scattering consists of the strong diffuse scattering in the vicinity of the former Bragg peak and a round spot limited by the critical angle $\theta_C$. The diffuse scattering at $\boldsymbol{Q} = \pm \boldsymbol{k}_s$ is maximal at $H \approx H_{c2}$ and strongly suppressed by increase of the field. According to Eq. (\ref{eq:1}) the spin-wave part of the scattering becomes narrower with further increase of the field and has vanished at a certain $H_{off}$ above $H_{c2}$. Using Eq.\ref{eq:0},  we define this value  as
\begin{equation}\label{eq:1b}
\theta_{0}^{2}- \frac{\theta_{0}}{E_i}H_{off} = 0
\end{equation} 
and obtain $ H_{off} = \theta_0 E_i$.
To define the cut-off angle $\theta_C$, a measurement of the background intensity at $H > H_{off}$ was subtracted from the other scattering maps. To improve the statistics, the scattering intensity was azimuthally-averaged over the angular sector of 120 degrees with the center positioned at $\boldsymbol{Q} = \pm \boldsymbol{k}_s$, as shown in Fig.\ref{ris:fig2}. The resultant curve is shown in Fig.\ref{ris:fig2} for $T = 250$ K and $H = 0.25$ T. From the analysis of the $I$ versus $\theta-\theta_B$ plot the cut-off angle can be extracted.

Nevertheless, a sharp cut-off of the intensity was not observed due to the dumping of spin-waves. The expected step-like intensity profile is smeared into the smoothly decreasing curve. This smeared step-like edge of the measured intensity was fitted by the following sigmoid function, which captures the main features of the scattering:
\begin{equation}\label{eq:2}
I(\theta) = I_0\left\{\frac{1}{2}- \left( \frac{1}{\pi}\arctan\left[\frac{2\left(\theta - \theta_C\right)}{\delta}\right]\right)\right\}
\end{equation}
The position of the cut-off angle was determined as the center of the arctan function $\theta_C$. Its width $\delta$ is related to the spin-wave damping $\Gamma$.

\begin{figure}[h]
                \begin{minipage}{0.99\linewidth}
                \center{\includegraphics[width=1\linewidth]{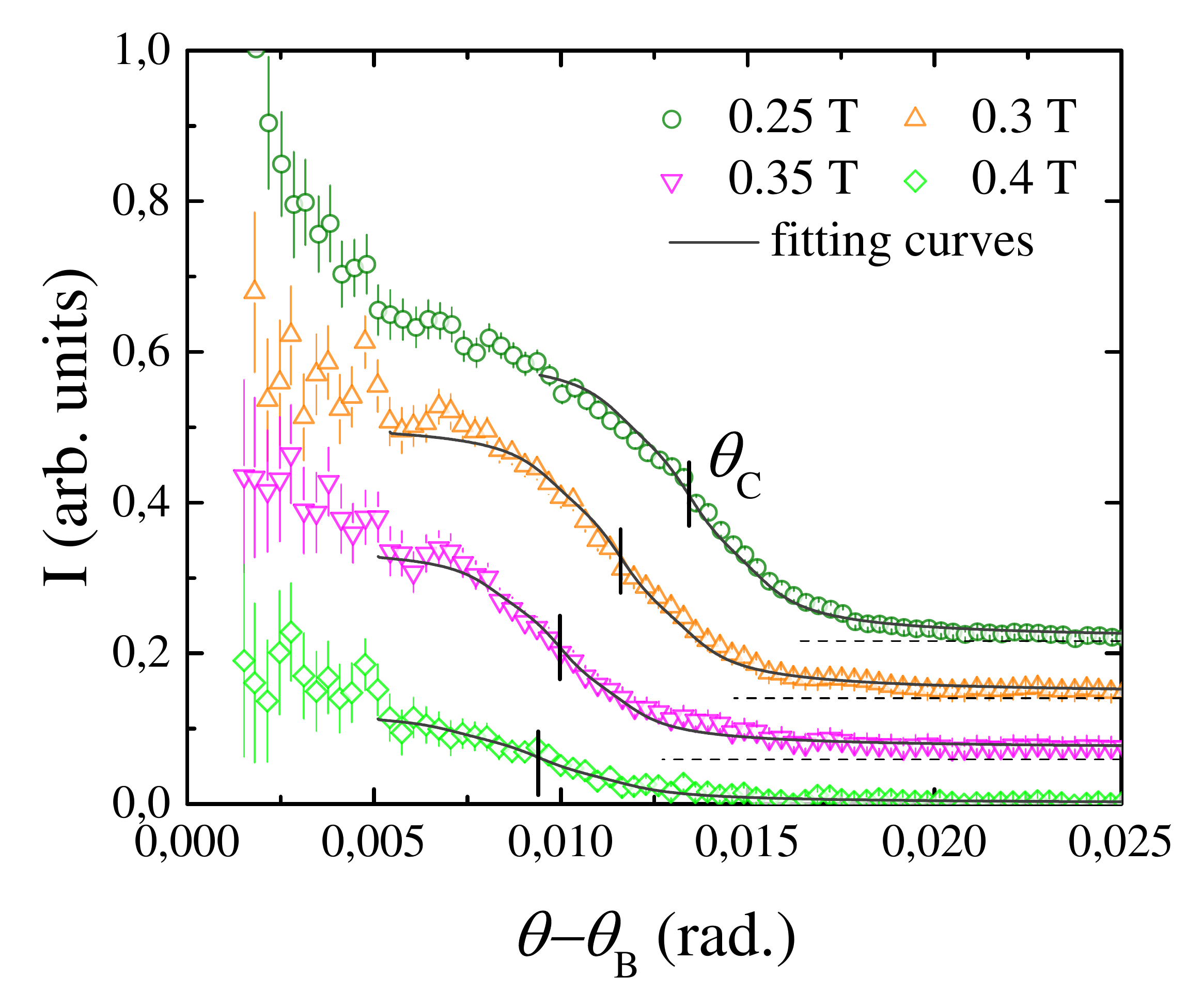}}
                \end{minipage}
                \caption{(color online). The azimuthally averaged intensities at $T = 250$ K for different magnetic field between $H = 0.25$ T and $H = 0.4$ T. The fitting functions are chosen analogously to Fig.\ref{ris:fig2}. The curves are shifted by a constant with respect to each other for clarity.}
                \label{ris:fig3}
\end{figure}

\begin{figure}[h]
        \begin{minipage}{0.99\linewidth}
        \center{\includegraphics[width=1\linewidth]{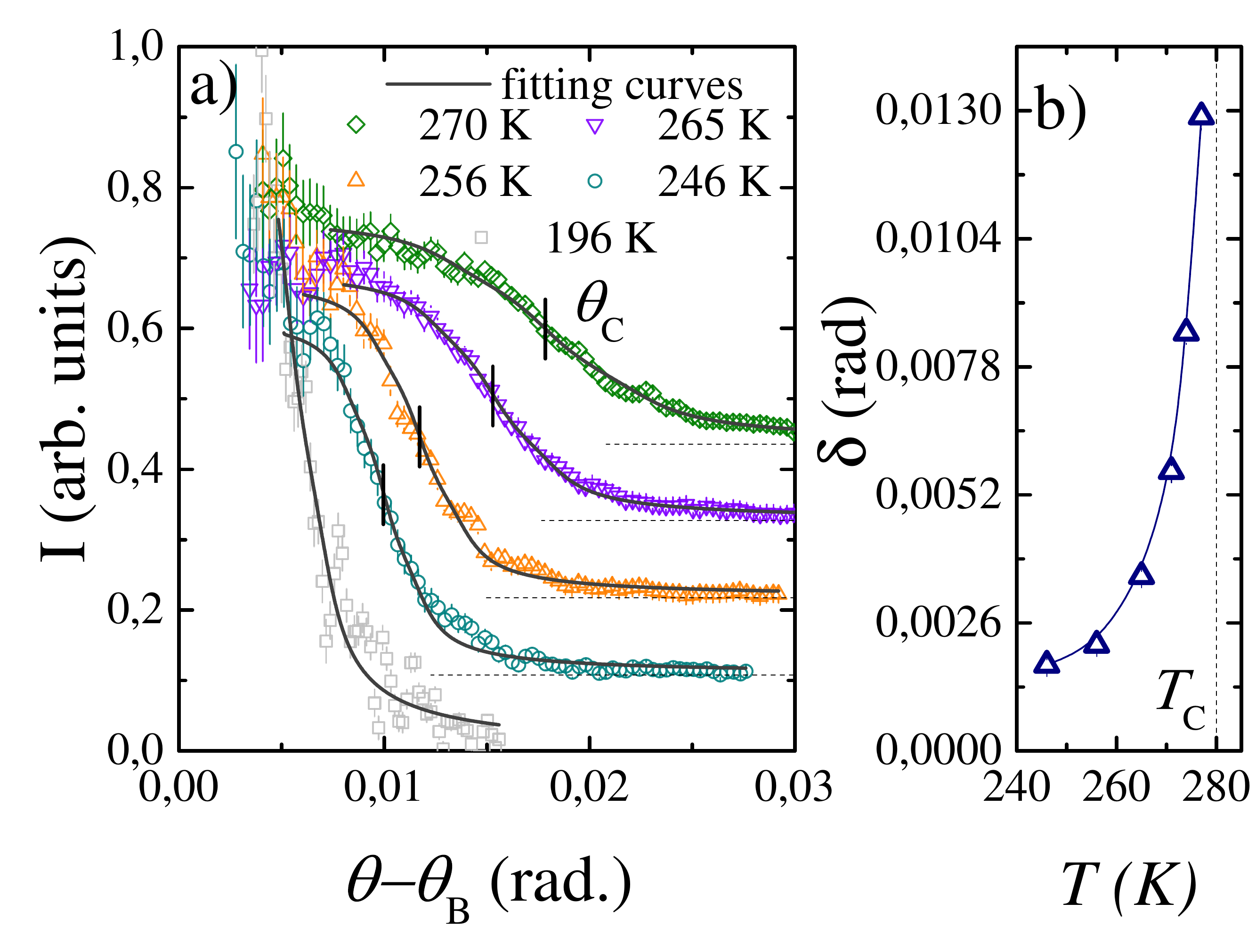}} 
        {\includegraphics[width=1\linewidth]{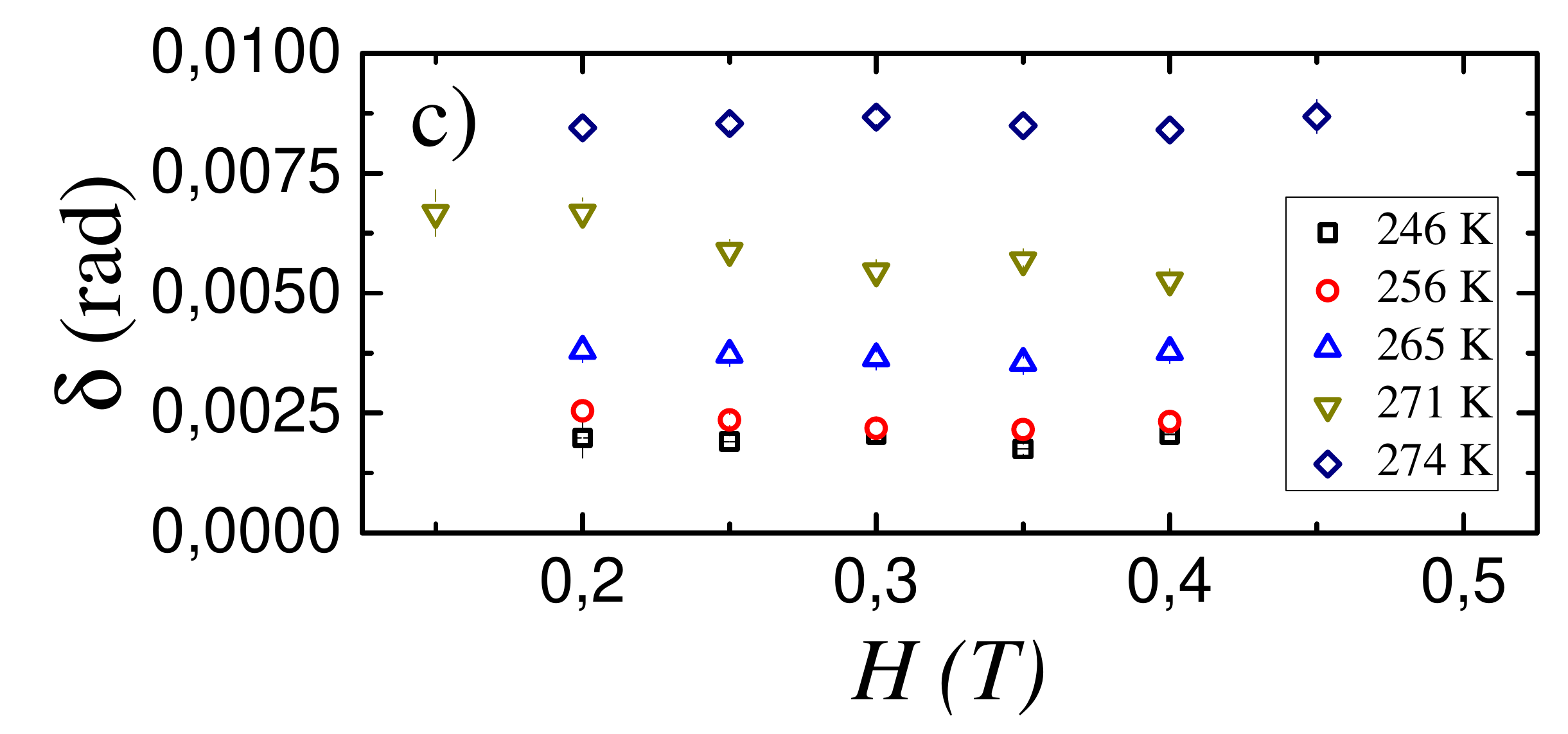}}
        \end{minipage}
        \caption{(color online). (a) The temperature evolution of the intensity profiles from $T = 196$ K to $T = 270$ K at $H = 0.35$ T. The curves are shifted by a constant with respect to each other for clarity. (b) The relative change of width of the step-like part $\delta$ for $H = 0.35$. (c) The relative change of width of the step-like part $\delta$ for $T = 246 K - 277 K$ between $H = 0.15$ - 0.45 T. }
        \label{ris:fig4}
\end{figure}

\begin{figure}[h]
	\begin{minipage}{0.99\linewidth}
		\center{\includegraphics[width=1\linewidth]{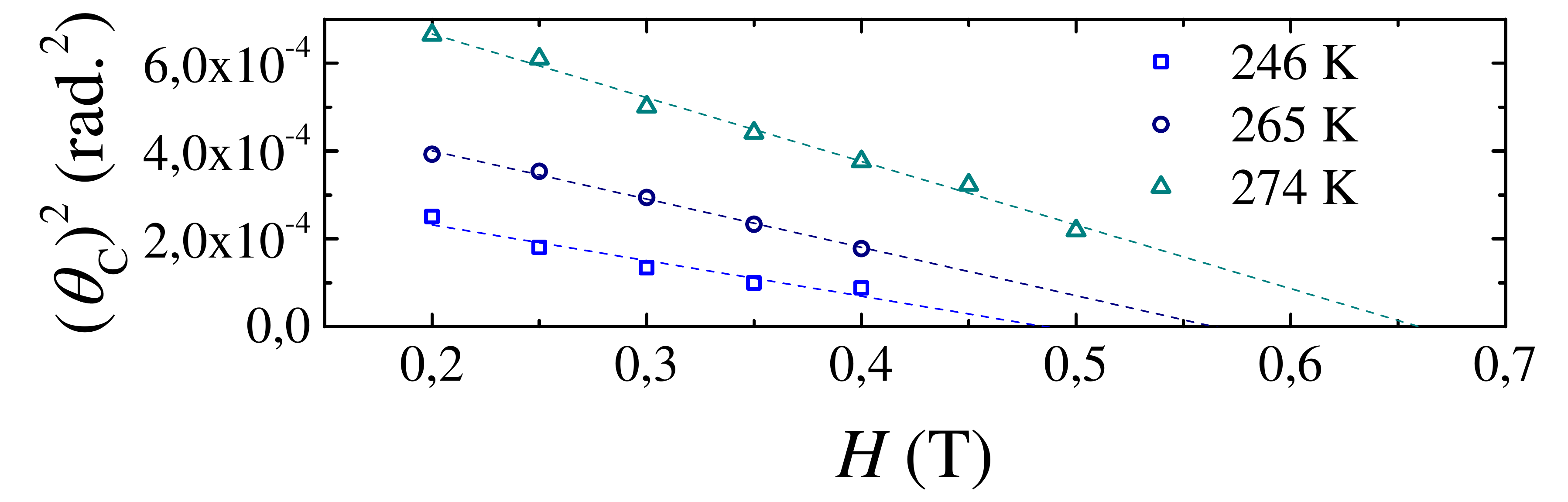}}
	\end{minipage}
	\caption{(color online). The field dependence of the square of the cut-off angle $\theta^{2}_{c}$ at $T$~=~246~K, 265~K and 274~K. The intersections between the linear fits and the $H$ axis determine $H_{off}$.}
	\label{ris:fig5}
\end{figure}

Fig.\ref{ris:fig3} shows the intensities as function of the angle $\theta-\theta_B$ at different values of the magnetic field for $T$~=~250~K together with the corresponding fitting curves. The curves show a decrease of the value of the cut-off angle together with a strong suppression of the scattering $\theta \approx \theta_B$ with increase of the field. The latter is related to the gap in the spin-wave spectrum increasing with the field, when the quasi elastic scattering with relatively low $\omega$ is forbidden. Consequently, the intensity being proportional to $1/\omega$ decreases at low $(\theta-\theta_B)$.

The intensity for $H$~=~0.35~T is plotted in Fig.\ref{ris:fig4} (a) for different temperatures between 196~K and 270~K including fitting functions. Two parameters $\theta_c$ and $\delta$ have been extracted from the fitting procedure. Fig.\ref{ris:fig4} (c) shows the field dependence of the dumping related parameter $\delta$ for different temperatures between 246~K and 277~K, which is nearly constant for each temperature.  Fig.\ref{ris:fig4} (b) shows the temperature dependence of $\delta$, $\delta$ increases drastically close to $T_C$ in accordance with theoretical expectations.  The linear field dependence of the squared cut-off angle $\theta_{C}^{2}$ against the magnetic field is shown in Fig.\ref{ris:fig5} for $T$~= 246, 265 and 274~K. 
We were not able to determine the cut-off angle for the temperatures below 225~K because of the intensive quasi-elastic scattering arising in the position of the former Bragg peak position.

\begin{figure}[h]
	\begin{minipage}{0.99\linewidth}
		\center{\includegraphics[width=1\linewidth]{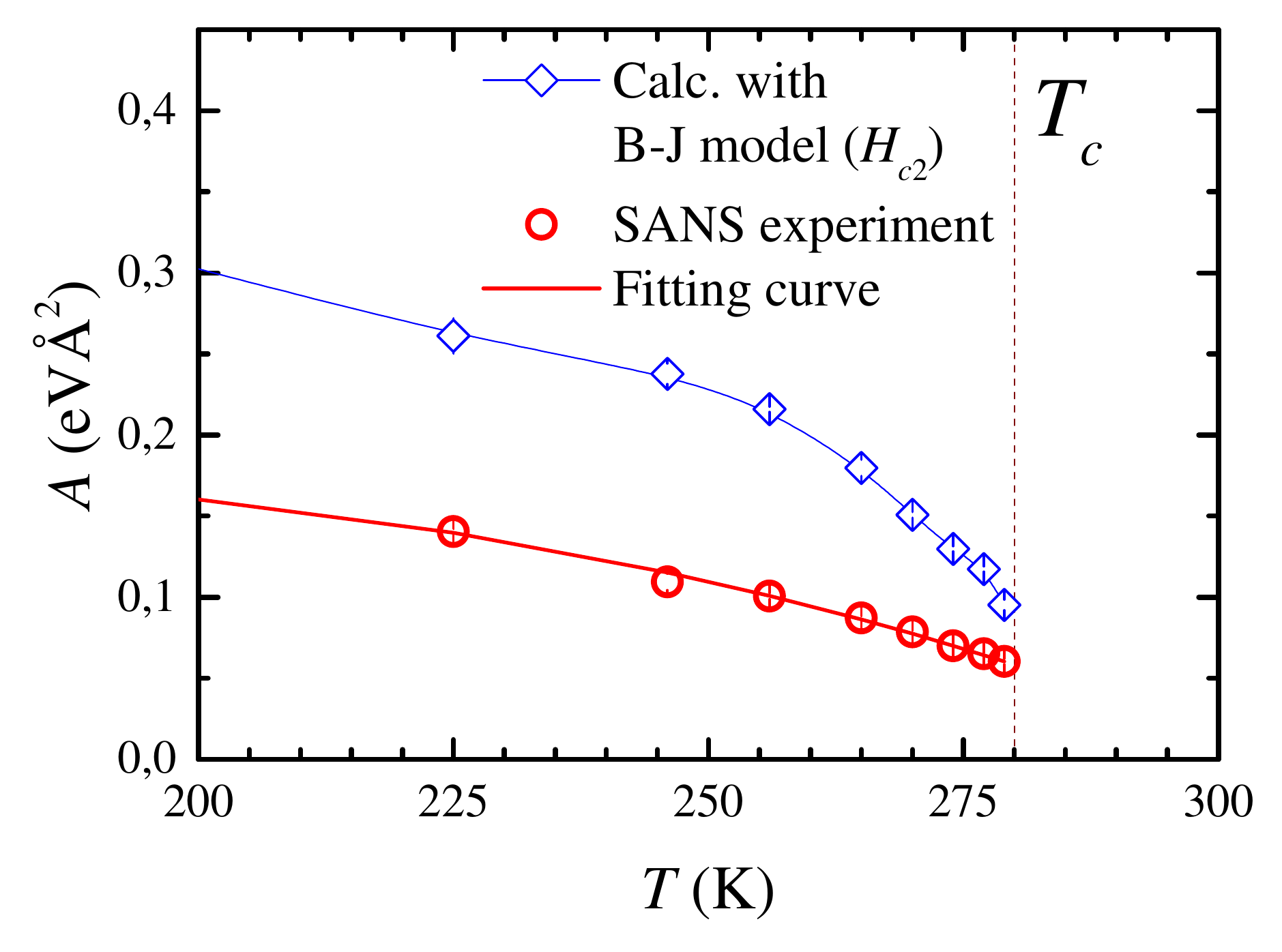}}
	\end{minipage}
	\caption{(color online). Temperature dependence of the spin-wave stiffness: open red circles measured by the cut-off angle with the corresponding fit, blue diamonds values estimated by the Bak-Jensen model using critical field $H_{c2}$. }
	\label{ris:fig6}
\end{figure}

The square of the cut-off angle depends linearly on the field in accordance with Eq. (\ref{eq:1}). With the help of Eq. (\ref{eq:1}) one can determine the value of the parameter $\theta_0$ and determine the spin-wave stiffness with high accuracy.
The spin-wave stiffness, obtained from the cut-off angle for different temperatures is shown in Fig.\ref{ris:fig6}. The measured temperature dependence was fitted by the power law: $A(T) = a\left(1-c\left(T/T_C\right)^{z}\right)$, parameters are found to be $a=(0.194 \pm 0.010)$ eV\AA$^2$, $c=0.70 \pm 0.01$ and $z=4.20 \pm 0.48$. 
The value of the stiffness $A$ does not tend to 0 as one would expect at $T_C$, for the second order phase transitions, but is finite $A \left(T_C\right) = a \cdot 0.3 = 0.058$~eV\AA$^2$. This fact clearly classifies the magnetic phase  transition in FeGe as being of the first order.

The spin-wave stiffness can also be estimated from the theory by Bak and Jensen \cite{Maleyev_PRB_2006} using the ratio relating the critical magnetic field $H_{c2}$ and the difference in the energies between the FP and helical states $g\mu_BH_{c2} = Ak_s$. The relation has been experimentally confirmed to be valid for MnSi in the whole temperature range below $T_C$ \cite{GrigorievPRB2015(R)}. The temperature dependence of the stiffness calculated in this model is also shown in Fig.\ref{ris:fig6}. The trend of the stiffness is the same including its decrease with temperature for the calculated and measured values. However, the magnitudes  strongly deviate one from another. As the matter of fact, the probed excitations show two times lower value than it was predicted from the value of the $H_{c2}$. The discrepancy between the experimental and calculated value of $A$ may be caused by the demagnetization effect within our polycrystalline samples, which has not been taken into account in determination of the critical field $H_{c2}$. Due to the higher magnetic moment and the polycrystalline nature of the used FeGe samples, this effect might play a bigger role for FeGe than for the previous investigated single crystalline MnSi \cite{GrigorievPRB2015(R)}.

\section{Conclusion}\label{sec:IV}
In conclusion, we have experimentally determined the spin-wave dynamic in the high temperature phase of the FeGe compound. We confirm the validity of the spin-wave dispersion relation for another helimagnet with DM interaction (Eq.(1)). Furthermore we demonstrated the ability of small-angle neutron scattering to measure the spin-wave dynamic in polycrystalline samples of DM helimagnets in the full-polarized state with acceptable statistics in reasonable time. The method allows the determination of the spin-wave dynamic in a broad temperature range and above all opens up complete new possibilities in the investigation of the parameter of the spin-wave dynamics in other representatives of DM helimagnets, which could be synthesized as powder only.
\\
\\
We thank Prof. S.V. Maleyev for valuable discussions and Prof. A. Schreyer for continuous interest and support.
The work was supported by the Russian Foundation of Basic Research (Grant No 14-22-01073,14-02-00001) and the special program of the Department of Physical Science, Russian Academy of Sciences.


%
\end{document}